\documentclass[3p]{elsarticle}
\usepackage{graphicx}
\usepackage{amsmath}
\usepackage{amssymb}
\biboptions{numbers,sort&compress}
\usepackage{graphicx}
\usepackage[usenames, dvipsnames]{color}
\usepackage{float}
\restylefloat{table}
\usepackage{lipsum}
\usepackage{mathtools}
\usepackage{cuted}
\usepackage{caption}
\usepackage{color}
\usepackage{array}
\usepackage{comment}
\usepackage{enumitem}
\makeatletter
\newcommand{\thickhline}{
    \noalign {\ifnum 0=`}\fi \hrule height 1pt
    \futurelet \reserved@a \@xhline
}
\newcolumntype{"}{@{\hskip\tabcolsep\vrule width 1pt\hskip\tabcolsep}}

\newtheorem{thm}{Theorem}
\newtheorem{lem}[thm]{Lemma}
\newdefinition{rmk}{Remark}
\newproof{pf}{Proof}
\newproof{pot}{Proof of Theorem \ref{thm2}}
\newdefinition{dfn}{Definition}
\newcommand{\pal}{\partial}
\newcommand{\al}{\alpha}
\makeatother
\journal{}
\begin{document}
\begin{frontmatter}
\title{Lie symmetry analysis of a class of time fractional nonlinear evolution systems}
\author[D]{Khongorzul Dorjgotov}
\author[O]{ Hiroyuki Ochiai}
\author[Z]{Uuganbayar Zunderiya}
\address[D]{Graduate School of Mathematics, Kyushu University, 744 Motooka, Fukuoka 819-0395, Japan}
\address[O]{Institute of Mathematics for Industry, Kyushu University, 744 Motooka, Fukuoka 819-0395, Japan}
\address[Z]{Department of Mathematics, National University of Mongolia, P.B. 507/38, Chingiltei 6, Ulaanbaatar 15141, Mongolia}
\begin{abstract}
We study a class of nonlinear evolution systems of time fractional partial differential equations using Lie symmetry analysis. We obtain not only infinitesimal symmetries but also a complete group classification and a classification of group invariant solutions of this class of systems. We find that the class of systems of differential equations studied is naturally divided into two cases on the basis of the type of a function that they contain. In each case, the dimension of the Lie algebra generated by the infinitesimal symmetries is greater than 2, and for this reason we present the structures and one-dimensional optimal systems of these Lie algebras. The reduced systems corresponding to the optimal systems are also obtained. Explicit group invariant solutions are found for particular cases.
\end{abstract}


\end{frontmatter}
\section{Introduction}
\label{intro}
Lie group analysis provides an efficient algorithmic approach for studying the  symmetry of ordinary and partial differential equations and for solving such equations \cite{Ovs,blu1,blu2,Ibr1,Ibr2,ibra,olver,Hydon}. Recently, the methods of symmetry analysis have been extended to solving fractional partial differential equations (FPDE)  \cite{Gaz2,Gaz3,Sah,Wang,Gaz1,gupta,Leo} and systems thereof \cite{huang,jeff,sing}. In recent years, there has been growing interest in FPDEs in fields of both pure and applied mathematics. In particular, FPDEs have been studied in the contexts of fractals, acoustics, control theory and signal processing. 

In this article, we consider the class of time fractional nonlinear systems of the following form:
\begin{equation}\label{1}
\begin{cases}
\frac{\pal^\al u}{\pal t^\al}=v_x,\\
\frac{\pal^\al v}{\pal t^\al}=b^2(u) u_x,
\end{cases}
\end{equation}
where $\alpha$ is a positive non-integer number and $b(u)$ is a non-constant, sufficiently differentiable function. Here, fractional differentiation is defined in the Riemann-Liouville manner: 
\begin{align}\label{defrld}
\frac{\pal^\al u(x,t)}{\pal t^\al}:=\left\{\begin{array}{ll}
\frac{\pal^n u}{\pal t^n},&\mbox{ for }\al\in\mathbb{N},\\
\frac{1}{\Gamma(n-\al)}\frac{\pal^n}{\pal t^n}\int_0^t\frac{u(x,s)}{(t-s)^{\al-n+1}}ds,&\mbox{ for } \al\in(n-1,n), \mbox{ with } n\in\mathbb{N}.
\end{array}
\right.
\end{align}

In \cite{huang}, the class of time fractional linear evolution systems 
\begin{equation}\label{qqqq1}
\begin{cases}
\frac{\partial^\al u}{\partial t^\al}=C^2(x)v_x,\\
\frac{\partial^\al v}{\partial t^\al}=u_x
\end{cases}
\end{equation}
where $C(x)$ is a sufficiently differentiable function and $\al$ is a positive non-integer number, is investigated using Lie symmetry analysis. Also, in \cite{jeff}, the nonlinear model of stationary transonic plane-parallel gas flows
\begin{equation}\label{qqqq2}
\begin{cases}
\frac{\partial^\al u}{\partial t^\al}=v_x,\\
\frac{\partial^\al v}{\partial t^\al}=-u u_x,
\end{cases}
\end{equation}
with $0<\al<1,$ was studied using Lie symmetry analysis. The Lie symmetries, some reduced systems of ODEs and some partial solutions of the system (\ref{qqqq2}) are obtained in \cite{jeff}. Substituting $\bar{u}(x,t)=-u(x,t)$ and $\bar{v}(x,t)=-v(x,t)$ into (\ref{qqqq2}), we obtain the following equivalent fractional system:
\begin{equation*}
\begin{cases}
\frac{\partial^\al \bar{u}}{\partial t^\al}=\bar{v}_x,\\
\frac{\partial^\al \bar{v}}{\partial t^\al}=\bar{u}\bar{u}_x.
\end{cases}
\end{equation*}
This corresponds to the particular case $b(u)=\sqrt{u}$ for the system given in (\ref{1}). It is thus seen that (\ref{1}) can be viewed as a nonlinear version of (\ref{qqqq1}) and a generalization of (\ref{qqqq2}) with respect to the above-mentioned substitution. Hence, the results of this paper generalize the results of \cite{jeff}.

The importance of finding exact solutions of (\ref{1}) lies in the fact that if $(u(x,t), v(x,t))$ solves (\ref{1}), then $u(x,t)$ solves the sequential equation
\begin{equation}\label{e2}
\frac{\pal^\al}{\pal t^\al}\frac{\pal^\al}{\pal t^\al}u=\left(b^2(u)u_x \right)_x.
\end{equation}
For example, in the case $\al=1,$ the equation (\ref{e2}) becomes the well-known nonlinear wave equation and in the case $\al=\frac{1}{2},$ the component $u(x,t)$ of the solutions of (\ref{1}) with $t>0$ is a solution to the nonlinear heat equation with source 
\begin{equation*}
u_t = \left(b^2(u)u_x\right)_x+\frac{1}{\sqrt{\pi}}\frac{g(x)}{\sqrt{t}},\qquad g(x) = \frac{1}{\sqrt{\pi}}\int_{0}^t\frac{u(x,\tau)}{\sqrt{t-\tau}}d\tau\biggr\vert_{t=0}, \quad t>0,
\end{equation*}
by virtue of the formula \cite{Pod}
\begin{equation*}
\frac{\pal^p}{\pal t^p}\frac{\pal^q}{\pal t^q}f(x,t)=\frac{\pal^{p+q}f(x,t)}{\pal t^{p+q}}-\sum_{j=1}^{n}\left[\frac{\pal^{q-j}f(x,t)}{\pal t^{q-j}}\right]\biggr\vert_{t=0}\frac{t^{m-p-j}}{\Gamma(1+m-p-j)},\mbox{ with } m-1\leq p<m,\quad n-1\leq q<n.
\end{equation*}
However, it should be noted that, in general, the Lie group of point transformations that leaves the system in (\ref{1}) invariant, does not necessarily correspond to a Lie group of point transformations that leaves the single equation in (\ref{e2}) invariant. 

We study the system given in (\ref{1}) using Lie  symmetry analysis. More explicitly, we present a complete group classification depending on the function $b(u)$ and describe the structure of Lie algebras generated by the infinitesimal symmetries of (\ref{1}). After obtaining the group classification of (\ref{1}), we proceed to finding optimal systems of Lie algebras and the reduced systems of ODEs. Using these optimal systems, we also classify the group invariant  solutions corresponding to the infinitesimal symmetries for $0<\al<1$. 

The organization of this paper is as follows. In Section 2, we present a simple introduction to the Lie symmetry analysis of systems of FPDEs and provide formulas useful in studying (\ref{1}). In Section 3, we carry out a complete group classification with respect to the function $b(u).$ In Section 4, we further investigate the structure of the corresponding Lie algebras of infinitesimal symmetries and determine the optimal systems. We also reduce (\ref{1}) to systems of fractional and non-fractional ODEs in accordance with these optimal systems. For particular cases, some explicit solutions are given in Section 5.
\section{Lie symmetry analysis for a system of fractional partial differential equations}\label{sec2}
To begin, we present the basic definitions and formulas needed to carry out the Lie symmetry analysis of a system of FPDEs. The general form of a system of time fractional PDEs with two independent variables $x$ and $t$ is as follows:
\begin{equation}\label{geneq}
\begin{cases}
\frac{\partial^\alpha u(x,t)}{\partial t^\alpha} = F_1(x,t,u,u_x,u_{xx},\ldots, v, v_x, v_{xx},\ldots),\\
\frac{\partial^\alpha v(x,t)}{\partial t^\alpha} = F_2(x,t,u,u_x,u_{xx},\ldots, v, v_x, v_{xx},\ldots),
\end{cases}
\end{equation}
where the subscripts denote partial derivatives and $\alpha$ is a positive real number. In the Lie symmetry analysis, the infinitesimal generator of (\ref{geneq}) is given by
\begin{equation*}
X=\xi\frac{\partial}{\partial x}+\tau\frac{\partial}{\partial t}+\mu\frac{\partial}{\partial u}+\phi\frac{\partial}{\partial v},
\end{equation*}
and the corresponding prolonged infinitesimal generator is 
\begin{equation}\label{gener}
\tilde{X}=X+\mu^{(\alpha)}\frac{\partial}{\partial u_{t^{\alpha}}}+\mu^{(1)}\frac{\partial}{\partial u_x}+\cdots
+\phi^{(\alpha)}\frac{\partial}{\partial v_{t^{\alpha}}}+\phi^{(1)}\frac{\partial}{\partial v_x}+\cdots,
\end{equation}
where $\tau,$ $\xi,$ $\mu$ and $\phi$ are infinitesimals and $\mu^{(\alpha)},$ $\mu^{(n)},$ $\phi^{(\alpha)}$ and $\phi^{(n)}$ $(n=1,2,\ldots)$ are extended infinitesimals. Explicitly, $\mu^{(n)}$ and $\phi^{(n)}$ are given by
\begin{align}\label{mun}
&\mu^{(1)}  =  D_x(\mu)-u_xD_x(\xi)-u_tD_x(\tau), \nonumber\\
&\mu^{(2)}  =  D_x(\mu^{(1)})-u_{xx}D_x(\xi)-u_{x t}D_x(\tau),\nonumber\\
&\vdots  \nonumber\\
&\phi^{(1)} = D_x(\phi)-v_xD_x(\xi)-v_tD_x(\tau),\nonumber\\
&\phi^{(2)} = D_x(\phi^{(1)})-v_{xx}D_x(\xi)-v_{x t}D_x(\tau),\nonumber\\
&\vdots 
\end{align}
where $D_x$ is the total derivative operator defined as
\begin{equation*}
D_x  :=  \frac{\partial}{\partial x}+u_x\frac{\partial}{\partial u}+u_{xx}\frac{\partial}{\partial u_x}+\cdots 
+v_x\frac{\partial}{\partial v}+v_{xx}\frac{\partial}{\partial v_x}+\cdots.
\end{equation*}
The $\alpha$th order extended infinitesimals have the following forms \cite{huang,sing}:
\begin{eqnarray}\label{mual1}
\mu^{(\alpha)} & = & D_t^\alpha (\mu)-\alpha D_t(\tau)\frac{\pal^\al u}{\pal t^\al}-\sum_{n=1}^\infty\binom{\al}{n}D_t^n(\xi)D_t^{\al-n}(u_x)-\sum_{n=1}^\infty\binom{\al}{n+1}D_t^{n+1}(\tau)D_t^{\al-n}(u),\nonumber\\
\phi^{(\alpha)} & = & D_t^\alpha (\phi)-\alpha D_t(\tau)\frac{\pal^\al v}{\pal t^\al}-\sum_{n=1}^\infty\binom{\al}{n}D_t^n(\xi)D_t^{\al-n}(v_x)-\sum_{n=1}^\infty\binom{\al}{n+1}D_t^{n+1}(\tau)D_t^{\al-n}(v).
\end{eqnarray}
Here, $D_t$ is the total derivative operator defined as
\begin{equation*}
D_t  :=  \frac{\partial}{\partial t}+u_t\frac{\partial}{\partial u}+u_{xt}\frac{\partial}{\partial u_x}+\cdots
+v_t\frac{\partial}{\partial v}+v_{xt}\frac{\partial}{\partial v_x}+\cdots.
\end{equation*}
Because the lower limit of the integral in (\ref{defrld}) is fixed, it should be invariant with respect to point transformations. We thus arrive at the initial condition 
\begin{equation}\label{incon}
\tau(x,t,u,v)|_{t=0}=0.
\end{equation}
We should note that the last three terms on the right-hand side of each equation in (\ref{mual1}) are already in factored forms with respect to the partial derivatives of $u$ and $v$. Hence, we need only consider the first terms, $D_t^\al(\mu)$ and $D_t^\al(\phi)$. 

In the following lemma, we present explicit forms of the extended infinitesimals $\mu^{(\al)},$ $\phi^{(\al)}$ that are readily computed.
\begin{lem}
The extended infinitesimals $\mu^{(\al)}$ and $\phi^{(\al)}$ in (\ref{mual1}) can be re-written as
\begin{eqnarray}\label{mual2}
\mu^{(\al)} & = & \frac{\pal^\al\mu}{\pal t^\al}-u\frac{\pal^\al\mu_u}{\pal t^\al}-v\frac{\pal^\al\mu_v}{\pal t^\al}+\left(\mu_u-\al D_t(\tau)\right)\frac{\pal^\al u}{\pal t^\al}+\mu_v\frac{\pal^\al v}{\pal t^\al}-\sum_{n=1}^\infty\binom{\al}{n}D_t^n(\xi)D_t^{\al-n}(u_x)\nonumber\\
& & +\sum_{n=1}^\infty\left[\binom{\al}{n}\frac{\pal^n\mu_u}{\pal t^n}-\binom{\al}{n+1}D_t^{n+1}(\tau)\right]D_t^{\al-n}(u)+\sum_{n=1}^\infty\binom{\al}{n}\frac{\pal^n\mu_v}{\pal t^n}D_t^{\al-n}(v)+\mu_1,
\end{eqnarray}
\begin{eqnarray}\label{phial}
\phi^{(\al)} & = & \frac{\pal^\al\phi}{\pal t^\al}-v\frac{\pal^\al\phi_v}{\pal t^\al}-u\frac{\pal^\al\phi_u}{\pal t^\al}+(\phi_v-\al D_t(\tau))\frac{\pal^\al v}{\pal t^\al}+\phi_u\frac{\pal^\al u}{\pal t^\al}-\sum_{n=1}^\infty\binom{\al}{n}D_t^n(\xi)D_t^{\al-n}(v_x)\nonumber\\
& & +\sum_{n=1}^\infty\left[\binom{\al}{n}\frac{\pal^n\phi_v}{\pal t^n}-\binom{\al}{n+1}D_t^{n+1}(\tau)\right]D_t^{\al-n}(v)+\sum_{n=1}^\infty\binom{\al}{n}\frac{\pal^n\phi_u}{\pal t^n}D_t^{\al-n}(u)+\phi_1,
\end{eqnarray}
where
\begin{eqnarray*}
\mu_1 & = & \sum_{n=2}^\infty\sum_{m_1+m_2=2}^n\sum_{\begin{subarray}
~k_1=0,..,m_1\\  
k_2=0,..,m_2\\  
k_1+k_2\geq 2
\end{subarray}}\sum_{r_1=0}^{k_1}\sum_{r_2=0}^{k_2}\binom{\al}{n}\binom{n}{m_1}\binom{n-m_1}{m_2}\binom{k_1}{r_1}\binom{k_2}{r_2} \frac{1}{k_1!k_2!}\frac{t^{n-\al}}{\Gamma(n+1-\al)}\\
&&\times(-u)^{r_1}(-v)^{r_2}\frac{\pal^{m_1}u^{k_1-r_1}}{\pal t^{m_1}}\frac{\pal^{m_2}v^{k_2-r_2}}{\pal t^{m_2}} \frac{\pal^{n-m_1-m_2+k_1+k_2}\mu}{\pal t^{n-m_1-m_2}\pal u^{k_1}\pal v^{k_2}},\nonumber
\end{eqnarray*}
and
\begin{eqnarray*}
\phi_1 &=& \sum_{n=2}^\infty\sum_{m_1+m_2=2}^n\sum_{\begin{subarray}
~k_1=0,..,m_1\\  
k_2=0,..,m_2\\  
k_1+k_2\geq 2
\end{subarray}}\sum_{r_1=0}^{k_1}\sum_{r_2=0}^{k_2}\binom{\al}{n}\binom{n}{m_1}
 \binom{n-m_1}{m_2}\binom{k_1}{r_1}\binom{k_2}{r_2} \frac{1}{k_1!k_2!}
\frac{t^{n-\al}}{\Gamma(n+1-\al)}\\
&&\times (-u)^{r_1}(-v)^{r_2}\frac{\pal^{m_1}u^{k_1-r_1}}{\pal t^{m_1}}\frac{\pal^{m_2}v^{k_2-r_2}}{\pal t^{m_2}}
\frac{\pal^{m-m_1-m_2+k_1+k_2}\phi}{\pal t^{n-m_1-m_2}\pal u^{k_1}\pal v^{k_2}}.
\end{eqnarray*}
\end{lem}
\begin{pf}
It is sufficient to prove the formula for $\mu^{(\al)}$. This can be done by carrying out an expansion of the first term $D_t^{\al}(\mu)$ in (\ref{mual1}) by applying  a generalized Leibniz rule and a generalized chain rule \cite{span, Pod}, as follows:
\begin{align}\label{k1k2}
D_t^\al(\mu) =& \sum_{n=0}^\infty\binom{\al}{n}\frac{t^{n-\al}}{\Gamma(n+1-\al)}D_t^n(\mu)\nonumber\\
=&\sum_{n=0}^\infty\sum_{m_1+m_2=0}^n\binom{\al}{n}\frac{t^{n-\al}}{\Gamma(n+1-\al)}\binom{n}{m_1}\binom{n-m_1}{m_2}\left.\left[\frac{\pal^{n-m_1-m_2}\pal^{m_1}\pal^{m_2}\mu(x,t,u(x,t_1),v(x,t_2))}{\pal t^{n-m_1-m_2}\pal t_1^{m_1}\pal t_2^{m_2}}\right]\right|_{\begin{subarray}{l}
t_1=t\\ 
t_2=t
\end{subarray}
}\nonumber\\
=&\sum_{n=0}^\infty\sum_{m_1+m_2=0}^n\sum_{k_1=0}^{m_1}\sum_{r_1=0}^{k_1}\sum_{k_2=0}^{m_2}\sum_{r_2=0}^{k_2}\binom{\al}{n}\frac{t^{n-\al}}{\Gamma(n+1-\al)}
\binom{n}{m_1}\binom{n-m_1}{m_2}\binom{k_1}{r_1}\binom{k_2}{r_2}\nonumber\frac{1}{k_1!k_2!}
\nonumber\\
&\times (-u)^{r_1}(-v)^{r_2}\frac{\pal^{m_1} u^{k_1-r_1}}{\pal t^{m_1}}\frac{\pal^{m_2} v^{k_2-r_2}}{\pal t^{m_2}}\frac{\pal ^{n-m_1-m_2+k_1+k_2}\mu}{\pal t^{n-m_1-m_2}\pal u^{k_1}\pal v^{k_2}}.
\end{align}
Because $\mu_1$ is equal to the partial sum obtained from (\ref{k1k2}) by retaining only the terms for which the sum of $k_1$ and $k_2$ is greater than 1, we need to examine only the case in which $k_1+k_2\leq 1$. All possible summations over values of $(m_1, m_2, k_1, k_2, r_1, r_2)$ satisfying this inequality can be divided into five cases, which we index by $i$. We write the resulting sum $D_t^\al(\mu)_{i)}$. These quantities are listed in the following table.
\begin{table}[H]
\centering
\begin{tabular}{|c|c|l|}
\hline
Subcase $i$ & $(m_1, m_2, k_1, k_2, r_1, r_2)$ & $D_t^\al(\mu)_{i)}$\\
\hline
1) & $(0,0,0,0,0,0)$ & $D_t^\al(\mu)_{1)}=\sum\limits_{n=0}^\infty\binom{\al}{n}D_t^{\al-n}1\cdot\frac{\pal^n\mu}{\pal t^n}=\frac{\pal^\al\mu}{\pal t^\al}$\\
2) & $(0,0,1,0,1,0)$ & $D_t^\al(\mu)_{2)}=\sum\limits_{n=0}^\infty\binom{\al}{n}D_t^{\al-n}1(-u)\frac{\pal^{n+1}\mu}{\pal t^n\pal u}=-u\frac{\pal^\al\mu_u}{\pal t^\al}$\\
3) & $(0,0,0,1,0,1)$ & $D_t^\al(\mu)_{3)}=-v\frac{\pal^\al\mu_v}{\pal t^\al}$\\
4) & $(m_1,0,1,0,0,0)$ & $D_t^\al(\mu)_{4)}=\mu_u\frac{\pal^\al u}{\pal t^\al}+\sum\limits_{n=1}^\infty\binom{\al}{n}\frac{\pal^n\mu_u}{\pal t^n}D_t^{\al-n}u$\\
5) & $(0,m_2,0,1,0,0)$ & $D_t^\al(\mu)_{5)}=\mu_v\frac{\pal^\al v}{\pal t^\al}+\sum\limits_{n=1}^\infty\binom{\al}{n}\frac{\pal^n\mu_v}{\pal t^n}D_t^{\al-n}v$\\
\hline
\end{tabular}
\end{table}

The explicit form of $\mu^{(\al)}$ given in the statement of the lemma can be obtained by substituting the sum of $D_t^\al(\mu)_{i)}$ in the above five cases  and $\mu_1$ into (\ref{mual1}). The formula for $\varphi^{(\al)}$ can be obtained similarly.
\qed
\end{pf}
Note: If $\mu^{(\al)}$ or $\phi^{(\al)}$ is linear in $u$ and $v,$ then $\mu_1=0$ and $\phi_1=0,$ respectively.

The infinitesimal invariance criterion in the Lie symmetry analysis for the system given in (\ref{geneq}) is
\begin{equation}\label{gende}
\begin{cases}
\left.\tilde{X}(u_{t^\alpha}-F_1(x,t,u,u_x,u_{xx},\ldots, v, v_x, v_{xx},\ldots))\right|_{(\ref{geneq})}=0,\\
\left.\tilde{X}(v_{t^\alpha}-F_2(x,t,u,u_x,u_{xx},\ldots, v, v_x, v_{xx},\ldots))\right|_{(\ref{geneq})}=0,
\end{cases}
\end{equation}
where $\tilde{X}$ is given by (\ref{gener}), (\ref{mun}), (\ref{mual2}) and (\ref{phial}). We are now ready to investigate the infinitesimal symmetries of the system in (\ref{1}) following the above-mentioned Lie symmetry analysis of the time fractional system.
\section{Lie symmetry analysis of the fractional nonlinear evolution system given in (\ref{1})}\label{sec3}
In this section, we study (\ref{1}) using the formulas obtained in the previous section. There are two cases regarding the symmetry group of (\ref{1}), as determined by the form of the function $b(u),$ one in which $b(u)$ possesses the form of a power function, and one in which it does not. The only difference between these cases is that in the former case, the symmetry group of (\ref{1}) possesses an additional symmetry that does not exist in the latter case. For each cases we obtain the infinitesimal symmetries. 

From (\ref{gende}), we obtain the following invariance criterion for (\ref{1}):
\begin{equation*}
\begin{cases}
\tilde{X}(u_{t^\al}-v_x)|_{(\ref{1})}=0,\\
\tilde{X}(v_{t^\al}-b^2(u)u_x)|_{(\ref{1})}=0.
\end{cases}
\end{equation*}
In explicit form, this is
\begin{equation}\label{10}
\begin{cases}
\left.\left(\mu^{(\al)}-\phi^{(1)}\right)\right|_{(\ref{1})}=0,\\
\left.(\phi^{(\al)}-2\mu b b' u_x-b^2\mu^{(1)})\right|_{(\ref{1})}=0.
\end{cases}
\end{equation}
From (\ref{10}), we obtain the following (overdetermined) system of determining equations by setting the coefficients of the linearly independent partial derivatives $D_t^{\al-n}u,$ $D_t^{\al-n}v,$ $D_t^{\al-n}u_x,$ $D_t^{\al-n}v_x,$ $v_x,$ $u_x,$ $v_t,$ $u_x v_t,$ $v_x v_t,$ $u_x v_x$ and $v_x^2$ equal to zero:
\begin{eqnarray*}
& & \binom{\al}{n}\frac{\pal^n\mu_u}{\pal t^n}-\binom{\al}{n+1}D_t^{n+1}\tau=0,\qquad n=1,2,\ldots,\\
& & \frac{\pal^n\mu_v}{\pal t^n}=0,\qquad n=1,2,\ldots,\\
& & D_t^n(\xi)=0,\qquad n=1,2,\ldots,\\
& & \mu_u-\al D_t(\tau)-\phi_v+\xi_x=0,\\
& & b^2\mu_v-\phi_u=0,\\
& & \frac{\pal^\al\mu}{\pal t^\al}-v\frac{\pal^\al\mu_v}{\pal t^\al}-u\frac{\pal^\al\mu_u}{\pal t^\al}-\phi_x+\mu_1=0,\\
& & \frac{\pal^n\phi_u}{\pal t^n}=0,\qquad n=1,2,\ldots,\\
& & \binom{\al}{n}\frac{\pal^n\phi_v}{\pal t^n}-\binom{\al}{n+1}D_t^{n+1}\tau=0,\qquad n=1,2,\ldots,\\
& & b^2\phi_v-\al b^2 D_t(\tau)-2b b'\mu-b^2\mu_u+b^2\xi_x=0,\\
& & \frac{\pal^\al\phi}{\pal t^\al}-v\frac{\pal^\al\phi_v}{\pal t^\al}-u\frac{\pal^\al\phi_u}{\pal t^\al}-b^2\mu_x+\phi_1=0,\\
& & \tau_x=\tau_u=\tau_v=0,\\
& & \xi_u=\xi_v=0.
\end{eqnarray*}
Analyzing the above overdetermined system with the initial condition (\ref{incon}), we are able to deduce the following infinitesimal symmetries:
\begin{itemize}
\item[Case 1.] This is the generic situation, which applies to all forms of $b(u),$ except $b(u)=ku^m$ (with $k,m\neq 0$). In this case, the infinitesimals are
\begin{equation*}
\tau=\frac{s_1}{\al}t,\quad \xi=s_1 x+s_2,\quad \mu=0,\quad \phi=s(t),
\end{equation*}
where $s_1$ and $s_2$ are arbitrary constants, and $s(t)$ is a solution of the equation $\frac{d^\al s(t)}{d t^\al}=0$. With these infinitesimals, we have the following infinitesimal symmetries:
\begin{equation*}
X_1=\frac{\pal}{\pal x},\quad X_2=s(t)\frac{\pal}{\pal v}, \quad X_3=x\frac{\pal}{\pal x}+\frac{t}{\al}\frac{\pal}{\pal t}.
\end{equation*}
\item[Case 2.] In the special case that $b(u)$ takes the form $ku^m$ (with $k,m\neq 0$), the infinitesimal are
\begin{equation*}
\tau=\frac{s_1}{\al}t,\quad \xi=(s_1+s_3) x+s_2,\quad \mu=\frac{s_3}{m}u,\quad \phi=\frac{(1+m)s_3}{m}v+s(t),
\end{equation*}
where $s_1,$ $s_2$ and $s_3$ are arbitrary constants, and $s(t)$ is a solution of the equation $\frac{d^\al s(t)}{d t^\al}=0$.
Thus, in this case, along with $X_1,$ $X_2$ and $X_3$ given above, there is the following additional symmetry:
\begin{equation*}
X_4=x\frac{\pal}{\pal x}+\frac{u}{m}\frac{\pal}{\pal u}+\frac{1+m}{m}v\frac{\pal}{\pal v}.
\end{equation*}
\end{itemize}

Because we now have a complete group classification of (\ref{1}), we are in a position to investigate the one-dimensional optimal systems of Lie algebras of its infinitesimal symmetries and the classification of group invariant solutions. However, before moving on to the next section, we note that the solution of the equation $\frac{d^\al s(t)}{d t^\al}=0$ is
\begin{equation*}
s(t)=c_1 t^{\al-1}+c_2 t^{\al-2}+\cdots +c_n t^{\al-n},
\end{equation*}
where $c_i$ are arbitrary constants and $n$ is a positive integer satisfying $n-1<\alpha<n$. 
\section{Optimal systems and the classification of invariant solutions}
An infinitesimal symmetry of a system of FPDEs generates a one-parameter group $G_a$ of transformations. The group $G_a$
maps any solution of the system in (\ref{geneq}) to a solution of the same system. An invariant  solution of (\ref{geneq}) is a solution that remains unaltered under any transformation in the group $G_a$. The solution $(u(x,t),v(x,t))$ of (\ref{geneq}) is an invariant of the group $G_a$  with the infinitesimal symmetry $X=\xi\frac{\pal}{\pal x}+\tau\frac{\pal}{\pal t}+\mu\frac{\pal}{\pal u}+\phi\frac{\pal}{\pal v}$ if and only if it solves the invariance surface condition \cite{ibra}
\begin{equation*}
\begin{cases}
\xi\frac{\pal u}{\pal x}+\tau\frac{\pal u}{\pal t}-\mu=0,\\
\xi\frac{\pal v}{\pal x}+\tau\frac{\pal v}{\pal t}-\phi=0.
\end{cases}
\end{equation*}

In this section, we classify the group invariant solutions of (\ref{1}) corresponding to infinitesimal symmetries for the case $0<\al<1$. The invariant solutions of (\ref{1}) corresponding to any infinitesimal symmetry can be obtained using Lie symmetry transformations applied to the invariant solutions corresponding to the infinitesimal symmetries of any optimal system of one-dimensional subalgebras of infinitesimal symmetries \cite{olver}. The optimal systems of low-dimensional Lie algebras are determined in \cite{pat}. For this reason, we need only to describe the invariant solutions corresponding to the infinitesimal symmetries of the optimal system.  More explicitly, we express the  invariant solutions as solutions of  reduced systems of ODEs. We choose optimal systems, which lead us to simpler reduced systems, by using the results given in \cite{pat}. In the following two subsections, we determine the optimal systems and corresponding reduced systems for Cases 1 and 2 specified above.
\subsection{Case 1.} 
From the discussion above, we know that in this case, for $\al$ satisfying $0<\al<1,$ the system in (\ref{1}) possesses the following infinitesimal symmetries:
\begin{equation*}
X_1=\frac{\pal}{\pal x},\quad X_2=t^{\alpha-1}\frac{\pal}{\pal v},\quad X_3=x\frac{\pal}{\pal x}+\frac{t}{\alpha}\frac{\pal}{\pal t}.
\end{equation*}
The commutator table for the Lie algebra generated by these infinitesimal symmetries is given below (where $i$ and $j$ index the row and column).
\begin{table}[H]\label{tab1}
\centering
\begin{tabular}{|c"c|c|c|} 
 \hline
 $[X_i, X_j]$ & $X_1$ & $X_2$ & $X_3$ \\ 
 \thickhline
 $X_1$ & 0 & 0 & $X_1$ \\ 
 \hline
 $X_2$ & 0 & 0 & $\frac{1-\alpha}{\alpha}X_2$ \\
 \hline
 $X_3$ & $-X_1$ & $-\frac{1-\alpha}{\alpha}X_2$ & 0\\
 \hline
\end{tabular}
\caption{Commutator table for Case 1.}
\end{table}
We see from this table that the Lie algebra in this case is identical to the Lie algebra $A_{3,5}$ given in \cite{pat}. Thus, the one-dimensional optimal system of the Lie algebra generated by $X_1$, $X_2$ and $X_3$ is that obtained in \cite{pat}, 
\begin{equation*}
U_1=X_1+aX_2\quad (\mbox{ with } a=0, 1, -1), \quad U_2=X_3, \quad U_3=X_2.
\end{equation*}
Then, using the standard characteristic method, we obtain the invariant solutions and reduced systems of ODEs of (\ref{1}) corresponding to each symmetry $U_j$. These are given below.
\begin{table}[H]\label{tab2}
\centering
\begin{tabular}{|c|l|l|}
    \hline
    $U_j$ & Invariant solutions $(u_j(x,t), v_j(x,t))$ & Reduced systems of ODEs  \\ 
    \thickhline
    $U_1$ & $\begin{array}{lcl }
    \begin{cases}
u(x,t) = \varphi(t),\\
v(x,t) = \psi(t)+a t^{\alpha-1}x,
\end{cases}
\end{array} $
 & $ \begin{array}{lcl}
 \begin{cases}
\frac{d^{\al} \varphi}{d t^\al} = a t^{\alpha-1},\\
\frac{d^{\al} \psi}{d t^\al} = 0,
\end{cases}
a=0, 1, -1
\end{array}
$\\\hline 
$U_2$ & $\begin{array}{lcl}
\begin{cases}
u(x,t)=\varphi(z),\\
v(x,t)=\psi(z),
\end{cases}
\end{array}$
 with $z=t x^{-\frac{1}{\alpha}}$
& $\begin{array}{lcl}
\begin{cases}
\frac{d^\al\varphi}{d z^\al}=-\frac{1}{\al}z\psi',\\
\frac{d^\al\psi}{d z^\al}=-\frac{1}{\al}z b^2(\varphi)\varphi',
\end{cases}
\end{array}$\\\hline
$U_3$ & \mbox{There are no invariant solutions.} & \\ \hline
  \end{tabular}
\caption{ The optimal systems and reduced systems of (\ref{1}) for Case 1}
\end{table}  
We consider the reduced systems of ODEs corresponding to $U_1$ in subsequent sections. The reduced system of ODEs corresponding to $U_2$ depends on $b(u)$, and for this reason, we do not solve it.
\subsection{Case 2.} 
To obtain the optimal systems for this case, here we construct both the commutator and adjoint tables for the Lie algebras of infinitesimal symmetries. The adjoint table is given by
\begin{equation*}
Ad(e^{(\varepsilon Y_i)})Y_j
=Y_j-\varepsilon [ Y_i,Y_j]+\frac{\varepsilon^2}{2}[Y_i,[Y_i,Y_j]]-\dots,\mbox{ where } \varepsilon\in\mathbb{R}.
\end{equation*}
In the present case, for $\al$ satisfying $0<\al<1$, the system in (\ref{1}) becomes
\begin{eqnarray}\label{15}
\begin{cases}
\frac{\pal^\alpha u}{\pal t^\alpha} = v_x,\\
\frac{\pal^\alpha v}{\pal t^\alpha} = k^2 u^{2m} u_x,
\end{cases}
\end{eqnarray}
and the corresponding infinitesimal symmetries are 
\begin{equation*}
X_1=\frac{\pal}{\pal x},\quad X_2=t^{\alpha-1}\frac{\pal}{\pal v},\quad X_3=x\frac{\pal}{\pal x}+\frac{t}{\alpha}\frac{\pal}{\pal t},\quad
X_4=x\frac{\pal}{\pal x}+\frac{u}{m}\frac{\pal}{\pal u}+\frac{m+1}{m}v\frac{\pal}{\pal v}.
\end{equation*}

The optimal systems of a given Lie algebra depend on the structure of that Lie algebra. Because the structure of the Lie algebra generated by the above $X_1,$ $X_2,$ $X_3$ and $X_4$ depends on the parameters $m$ and $\al$, we study (\ref{15}) in two subcases characterized by the relations $2m\alpha+\alpha-m\neq 0$ and $2m\alpha+\alpha-m=0$. 
\subsubsection*{Case 2.1.}
Let us consider the case $2m\alpha+\alpha-m\neq 0$. Then, for the Lie algebra generated by the symmetries $X_i$ $(i=1,\ldots,4),$ we choose a new basis  $Y_i$ $(i=1,\ldots,4)$ such that the Lie algebra consists of the  direct sum of two subalgebras $L_1$ and $L_2$, where $L_1$ is generated by $Y_1$ and $Y_2,$ and $L_2$ is generated by $Y_3$ and $Y_4$. We choose this new basis as follows:
\begin{equation*}
Y_1=-\frac{(m+1)\alpha}{2m\alpha+\alpha-m}X_3-\frac{m(\alpha-1)}{2m\alpha+\alpha-m}X_4,
\quad Y_2=-X_1,\quad Y_3=\frac{m\alpha}{2m\alpha+\alpha-m}(X_3-X_4),\quad Y_4=-X_2.
\end{equation*}
The commutator and adjoint tables for $Y_i$ $(i=1,\ldots,4)$ are given below (where $i$ and $j$ index the row and column).
\begin{table}[H]\label{tab3}
\centering
\begin{tabular}{|c"c|c"c|c|}
\hline
$[Y_i, Y_j]$ & $Y_1$ & $Y_2$ & $Y_3$ & $Y_4$\\
\thickhline
$Y_1$ & 0 & $Y_2$ & 0 & 0\\
\hline
$Y_2$ & $-Y_2$ & 0 & 0 & 0\\
\thickhline
$Y_3$ & 0 & 0 & 0 & $Y_4$\\
\hline
$Y_4$ & 0 & 0 & $-Y_4$ & 0\\
\hline
\end{tabular}
\caption{Commutator table for Case 2.1.}
\end{table}
\begin{table}[H]\label{tab4}
\centering
\begin{tabular}{|c"c|c|c|c|}
\hline
$Ad(e^{(\varepsilon Y_i)})Y_j$ & $Y_1$ & $Y_2$ & $Y_3$ & $Y_4$\\
\thickhline
$Y_1$ & $Y_1$ & $e^{-\epsilon Y_2}$ & $Y_3$ & $Y_4$\\
\hline
$Y_2$ & $Y_1+\epsilon Y_2$ & $Y_2$ & $Y_3$ & $Y_4$\\
\hline
$Y_3$ & $Y_1$ & $Y_2$ & $Y_3$ & $e^{-\epsilon Y_4}$\\
\hline
$Y_4$ & $Y_1$ & $Y_2$ & $Y_3+\epsilon Y_4$ & $Y_4$\\
\hline
\end{tabular}
\caption{Adjoint table for Case 2.1.}
\end{table}
We choose the following optimal system for the Lie algebra generated by $Y_i$ $(i=1,\dots,4)$ to simplify the reduced systems of FODEs:
\begin{eqnarray*}
U_{1} = & Y_2+aY_4 & =- X_1-aX_2,\quad a=0,1,-1,\\
U_{2} = & Y_2+\frac{(2m\alpha+\alpha-m)a}{m\alpha}Y_3 & =-X_1+aX_3-aX_4,\quad a=1,-1,\\
U_{3} = & (2m\alpha+\alpha-m)Y_1+aY_4 & =-aX_2-(m+1)\al X_3-m(\al-1)X_4,\quad a=1,-1,  \\ 
U_4 = & Y_1+\frac{(2m\alpha+\alpha-m)a-m\al+m}{m\alpha}Y_3 & =(a-1)X_3-aX_4,\quad a\in \mathbb{R},\\
U_5 = & \frac{2m\alpha+\alpha-m}{m\alpha}Y_3 & =X_3-X_4, \\ 
U_{6} = & Y_4 & =-X_2.
\end{eqnarray*}
\begin{rmk}
We see from Table 3 that this Lie algebra is identical to the Lie algebra $2A_2$ given in \cite{pat}. If we act on $U_2$ and $U_3$ with $Ad(e^{\epsilon Y_1})$ and $Ad(e^{\epsilon Y_3}),$ respectively, with suitable $\epsilon,$ we obtain the equivalences $U_2\sim Y_1+a Y_3$ and $U_3\sim Y_1+a Y_4$ $(a=\pm 1).$ This demonstrates the correspondence between the optimal system chosen here and the optimal system in \cite{pat}.
\end{rmk}

In Tables 5 and 6, we display the similarity variables $z_j,$ invariant solutions $(u_j(x,t), v_j(x,t))$ expressed as solutions $(\varphi(z),\psi(z))$ of the reduced systems of ODEs, and the reduced system of ODEs corresponding to $U_j$ in the optimal system. Note that due to the divergence of the integral in the definition (\ref{defrld}) of the Riemann-Liouville derivative, $\frac{d^\al}{dt^\alpha}(t^p)$ is not defined for $p\leq -1$ \cite{span}. For this reason, here we need an additional assumption, which is expressed in Table 5, regarding invariant solutions corresponding to the symmetry $U_5$.
\begin{table}[H]\label{tab5}
\centering
\begin{tabular}{|c|l|l|}
  \hline
    $U_j$ & $z_j$ & Invariant solutions $(u_j(x,t), v_j(x,t))$ \\ 
    \thickhline
    $U_{1}$ & $t$ & $\begin{array}{lcl }
    \begin{cases}
u(x,t) = \varphi(t),\\
v(x,t) = \psi(t)+axt^{\alpha-1},
\end{cases}a=0,1,-1
\end{array} $\\
\hline
$U_{2}$ & $t\exp\left(\frac{a}{\alpha}x\right)$ & $\begin{array}{lcl}
\begin{cases}
u(x,t)=\exp\left(\frac{a}{m}x\right)\varphi(z),\\
v(x,t)=a\exp\left(\frac{(m+1)a}{m}x\right)\psi(z),
\end{cases}
a=1,-1
\end{array}$\\ 
\hline
$U_{3}$ & $tx^{-\frac{m+1}{2m\alpha+\alpha-m}}$ & $\begin{array}{lcl}
\begin{array}{l}
\begin{cases}
u(x,t)=x^{\frac{\alpha-1}{2m\alpha+\alpha-m}}\varphi(z),\\
v(x,t)=x^{\frac{(m+1)(\alpha-1)}{2m\alpha+\alpha-m}}\psi(z)+\frac{a}{2m\alpha+\alpha-m}t^{\alpha-1}\ln(x),
\end{cases}
a=1,-1
\end{array}
\end{array}$\\ 
\hline
$U_4$ & $t x^{\frac{a-1}{\alpha}}$ & $\begin{array}{lcl}
\begin{cases}
u(x,t)=x^{\frac{a}{m}}\varphi(z),\\
v(x,t)=x^{\frac{(m+1)a}{m}}\psi(z),
\end{cases}
a\in \mathbb{R}
\end{array}$ \\ \hline
$U_{5}$ & $x$ & $\begin{array}{lcl}
\begin{cases}
u(x,t)=t^{-\frac{\alpha}{m}}\varphi(x),\\
v(x,t)=t^{-\frac{(m+1)\alpha}{m}}\psi(x),
\end{cases}
m<0 \mbox{ or } m>\frac{\al}{1-\al}
\end{array}$ \\ \hline
$U_{6}$ &  & There are no invariant solutions.\\ 
\hline
\end{tabular}
\caption{Similarity variables $z_j$ and invariant solutions $(u_j, v_j)$ for Case 2.1.}
\end{table}
\begin{table}[H]\label{tab6}
\centering
\begin{tabular}{|c|l|}
    \hline
    $U_j$ & Reduced system of ODEs \\ 
    \thickhline
   $U_{1}$ & $\begin{array}{lcl}
\begin{cases}
\frac{d^\alpha\varphi(t)}{d t^\alpha}=at^{\alpha-1},\\
\frac{d^\alpha\psi(t)}{d t^\alpha}=0,
\end{cases}
a=0,1,-1
\end{array}$\\ \hline
$U_{2}$ & $\begin{array}{lcl}
\begin{cases}
\frac{d^\alpha\varphi}{d z^\alpha}=\left(\frac{m+1}{m}\psi+\frac{1}{\alpha} z \psi'\right),\\
\frac{d^\alpha\psi}{d z^\alpha}=k^2\varphi^{2m}\left(\frac{1}{m}\varphi+\frac{1}{\alpha} z \varphi'\right),
\end{cases}
\end{array}$\\ \hline
$U_{3}$ & $\begin{array}{lcl}
\begin{cases}
\frac{d^\alpha\varphi}{d z^\alpha}=\frac{m+1}{2m\alpha+\alpha-m}\left((\alpha-1)\psi-z\psi'\right)+\frac{a}{2m\alpha+\alpha-m}z^{\alpha-1},\\
\frac{d^\alpha\psi}{d z^\alpha}=\frac{k^2}{2m\alpha+\alpha-m}\varphi^{2m}\left((\alpha-1)\varphi-(m+1)z\varphi'\right),
\end{cases}
a=1,-1
\end{array}$\\ \hline
$U_4$ & $\begin{array}{lcl}
\begin{cases}
\frac{d^\alpha\varphi}{d z^\alpha}=\frac{(m+1)a}{m}\psi+\frac{a-1}{\alpha}z\psi',\\
\frac{d^\alpha\psi}{d z^\alpha}=k^2\varphi^{2m}\left(\frac{a}{m}\varphi+\frac{a-1}{\alpha}z\varphi'\right),
\end{cases}
a\in \mathbb{R}
\end{array}$\\ \hline
$U_{5}$ & $\begin{array}{lcl}
\begin{cases}
\psi'(x)=\frac{\Gamma\left(1-\frac{\alpha}{m}\right)}{\Gamma\left(1-\frac{(m+1)\alpha}{m}\right)}\varphi(x),\\
k^2\varphi^{2m}\varphi'(x)=\frac{\Gamma\left(1-\frac{(m+1)\alpha}{m}\right)}{\Gamma\left(1-\frac{(2m+1)\alpha}{m}\right)}\psi(x),
\end{cases}
m<0 \mbox{ or } m>\frac{\al}{1-\al}
\end{array}$\\ \hline
\end{tabular}
\caption{Reduced systems of ODEs for Case 2.1.}
\end{table}
\subsubsection*{Case 2.2.} 
Next, let us consider the case $2m\alpha+\alpha-m=0$. In this case, we obtain $m=\frac{\alpha}{1-2\alpha}$ with $\alpha\neq \frac{1}{2}$. Then, for the Lie algebra generated by the symmetries $X_i$ $(i=1,\ldots,4),$ we again choose a new basis  $Y_i$ $(i=1,\ldots,4)$ such that the Lie algebra consists of the  direct sum of two subalgebras $L_1$ and $L_2$. However, in this case $L_1$ is generated by $Y_1,$ $Y_2$ and $Y_3,$ and $L_2$ is generated by $Y_4$. The new basis is
\begin{equation*}
Y_1=X_1,\quad Y_2=X_2,\quad Y_3=X_4,\quad Y_4=X_4-X_3.
\end{equation*}
The commutator and adjoint tables for $Y_i$ $(i=1,\ldots,4)$ are given below:  \begin{table}[H]\label{tab7}
\centering
\begin{tabular}{|c"c|c|c"c|}
\hline
$[Y_i, Y_j]$ & $Y_1$ & $Y_2$ & $Y_3$ & $Y_4$\\
\thickhline
$Y_1$ & 0 & 0 & $Y_1$ & 0\\
\hline
$Y_2$ & 0 & 0 & $\frac{1-\al}{\al}Y_2$ & 0\\
\hline
$Y_3$ & $-Y_1$ & $-\frac{1-\al}{\al}Y_2$ & 0 & 0\\
\thickhline
$Y_4$ & 0 & 0 & 0 & 0\\
\hline
\end{tabular}
\caption{Commutator table for Case 2.2.}
\end{table}
\begin{table}[H]\label{tab8}
\centering
\begin{tabular}{|c"c|c|c|c|}
\hline
$Ad(e^{(\varepsilon Y_i)})Y_j$ & $Y_1$ & $Y_2$ & $Y_3$ & $Y_4$\\
\thickhline
$Y_1$ & $Y_1$ & $Y_2$ & $Y_3-\varepsilon Y_1$ & $Y_4$\\
\hline
$Y_2$ & $Y_1$ & $Y_2$ & $Y_3-\frac{1-\alpha}{\al}\varepsilon Y_2$ & $Y_4$\\
\hline
$Y_3$ & $e^\varepsilon Y_1$ & $e^{\frac{1-\al}{\al}\varepsilon}Y_2$ & $Y_3$ & $Y_4$\\
\hline
$Y_4$ & $Y_1$ & $Y_2$ & $Y_3$ & $Y_4$\\
\hline
\end{tabular}
\caption{Adjoint table for Case 2.2.}       
\end{table}
We choose the following optimal system in the new and original bases:
\begin{eqnarray*}
U_{1}&=&Y_1+aY_2=X_1+aX_2,\quad a=0,1,-1,\\
U_{2}&=&Y_1+a_1 Y_2+a_2Y_4\\ 
& =&X_1+a_1 X_2-a_2X_3+a_2 X_4,\quad (a_1,a_2)\in\{(a,\pm 1)|a\in \mathbb{R}\},\\
U_3&=&Y_3+(a-1) Y_4 = (1-a)X_3+aX_4,\quad a\in \mathbb{R},\\
U_{4}&=&aY_2+Y_4 =aX_2-X_3+X_4,\quad a=0,1,-1,\\
U_5&=&Y_2  =X_2.
\end{eqnarray*}
We see from Table 7 that this Lie algebra is identical to the Lie algebra $A_{3,5}\oplus A_1$ given in \cite{pat}. Hence, considering the following equivalences regarding the actions of $Ad(e^{\epsilon Y_3})$ on $U_2,$ it is seen that
our optimal system of this Lie algebra corresponds bijectively to that given in \cite{pat}.
\begin{enumerate}[label=\alph*)]
\item For any $a>0$, there exists $b>0$ such that $Y_1+aY_2+Y_4\sim Y_1+Y_2+bY_4$,  
\item For any $a>0$, there exists $b<0$ such that $Y_1+aY_2-Y_4\sim Y_1+Y_2+bY_4$,
\item For any $a<0$, there exists $b>0$ such that $Y_1+aY_2+Y_4\sim Y_1-Y_2+bY_4$,  
\item For any $a<0$, there exists $b<0$ such that $Y_1+aY_2-Y_4\sim Y_1-Y_2+bY_4$. 
\end{enumerate}

In the following Table 9, we display the similarity variables $z_j$ and invariant solutions $(u_j(x,t),v_j(x,t)),$ which are expressed as solutions of reduced systems. Then, in Table 10, we present the reduced systems of ODEs corresponding to the above optimal system.
\begin{table}[H]
\centering
\begin{tabular}{|c|l|l|}
    \hline
    $U_j$ & $z_j$ & Invariant solutions $(u_j(x,t),v_j(x,t))$ \\     \thickhline
    $U_1$ & $t$ & $\begin{array}{lcl}
    \begin{cases}
u(x,t) = \varphi(t),\\
v(x,t) = \psi(t)+a xt^{\alpha-1},
\end{cases}
a=0,1,-1
\end{array} $
\\
\hline 
    $U_{2}$ & $t\exp(\frac{a_2}{\alpha}x)$ & 
$\begin{array}{lcl}
\begin{cases}
u(x,t)=\exp\left(\frac{a_2(1-2\alpha)}{\alpha}x\right)\varphi(z),\\
v(x,t)=a_2\exp\left(\frac{a_2(1-\alpha)}{\alpha}x\right)\psi(z)+a_1xt^{\alpha-1},
\end{cases}
(a_1,a_2)\in\{(a,\pm 1)|a\in \mathbb{R}\}
\end{array}$\\
\hline
$U_3$ & $t x^{\frac{a-1}{\alpha}}$ & $\begin{array}{lcl}
\begin{cases}
u(x,t)=x^\frac{a(1-2\alpha)}{\alpha}\varphi(z),\\
v(x,t)=x^\frac{a(1-\alpha)}{\alpha}\psi(z),
\end{cases}
a\in \mathbb{R}
\end{array}$\\ \hline
$U_{4}$ & $x$ & $\begin{array}{lcl}
\begin{cases}
u(x,t)=t^{2\alpha-1}\varphi(x),\\
v(x,t)=t^{\alpha-1}\psi(x)-a\alpha t^{\alpha-1}\ln(t),
\end{cases}
a=0,1,-1
\end{array}$\\ \hline
$U_5$ &  & There are no invariant solutions.\\ \hline
\end{tabular}
\caption{Similarity variables $z_j$ and invariant solutions $(u_j,v_j)$ for Case 2.2.}
\end{table}
\begin{table}[H]
\centering
\begin{tabular}{|c|l|}
    \hline
    $U_j$ & Reduced system of ODEs \\ 
    \thickhline
    $U_1$ & $ \begin{array}{lcl}
 \begin{cases}
\frac{d^\al \varphi(t)}{d t^\al} = at^{\alpha-1},\\
\frac{d^\al \psi(t)}{d t^\al} = 0,
\end{cases}
a=0,1,-1
\end{array}
$
\\
\hline 
    $U_{2}$ & $\begin{array}{lcl}
\begin{cases}
\frac{d^\alpha\varphi}{d z^\alpha}=\frac{1}{\alpha}\left((1-\alpha)\psi+z\psi'\right)+a_1z^{\alpha-1},\\
\frac{d^\alpha\psi}{d z^\alpha}=\frac{k^2}{\alpha} \varphi^{\frac{2\alpha}{1-2\alpha}}\left((1-2\alpha)\varphi+z\varphi'\right),
\end{cases} 
a_1\in \mathbb{R}
\end{array}$\\ \hline
$U_3$ & $\begin{array}{lcl}
\begin{cases}
\frac{d^\alpha\varphi}{d z^\alpha}=\frac{1}{\alpha}\left(a(1-\alpha)\psi+(a-1)z\psi'\right),\\
\frac{d^\alpha\psi}{d z^\alpha}= \frac{k^2}{\alpha} \varphi^{\frac{2\alpha}{1-2\alpha}}\left(a(1-2\alpha)\varphi+(a-1)z\varphi'\right),
\end{cases}
a\in \mathbb{R}
\end{array}$\\ \hline
$U_{4}$ & $\begin{array}{lcl}
\begin{cases}
\psi'(x)=\frac{\Gamma(2\alpha)}{\Gamma(\alpha)}\varphi(x),\\
k^2\varphi^\frac{2\alpha}{1-2\alpha}\varphi'(x)=-a\Gamma(\alpha+1),
\end{cases}
a=0,1,-1,
\end{array}$\\ \hline
\end{tabular}
\caption{Reduced systems of ODEs for Case 2.2.}
\end{table}
With the above results for the optimal systems in Cases 2.1 and 2.2, we arrive at the following three conclusions for the case $m=\frac{\al}{1-2\al}$.
\begin{enumerate}
\item The two invariant solutions corresponding to $U_1$ in Cases 2.1 and 2.2 coincide.
\item The invariant solutions corresponding to $U_3$ in Case 2.1 coincide with the invariant solutions corresponding to $U_4$ in Case 2.2.
\item The invariant solutions corresponding to $U_4$ in Case 2.1 coincide with the invariant solutions corresponding to $U_3$ in Case 2.2.
\end{enumerate}
Even though the elements of the optimal systems in Cases 2.1 and 2.2 correspond to each other, except for the element $U_2$, the reduced systems of ODEs in Cases 2.1 and 2.2 generally differ, possessing a relationship determined by the choice of the similary variable $z$.

In the next section, we derive several explicit invariant solutions of the fractional system in (\ref{1}) by solving the reduced systems of ODEs obtained in this section.
\section{Explicit solutions of the fractional nonlinear evolution system given in (\ref{1})}
In the general case, solving fractional order nonlinear systems of ODEs is a challenging problem. However, here we are able to derive several explicit solutions of the reduced systems of ODEs obtained in the previous section. Then, using these solutions, we  obtain several group invariant solutions of (\ref{1}).
\subsection{Reduced system of ODEs corresponding to $U_1$}
The reduced systems of ODEs corresponding to $U_1$ are essentially the same in all three cases. The following is the solution to each of these:
\begin{equation*}
\begin{cases}
\varphi(t) = \frac{a}{\Gamma(\al)}t^{2\al-1}+c_1 t^{\al-1},\\
\psi(t) = c_2 t^{\al-1},
\end{cases}
\end{equation*}
where $c_1$ and $c_2$ are arbitrary constants. With the above, we obtain the following for the invariant solution of (\ref{1}):
\begin{equation}
\begin{cases}
u(x,t)=\frac{a}{\Gamma(\al)}t^{2\al-1}+c_1 t^{\al-1},\\
v(x,t)=(c_2+a x)t^{\al-1}.
\end{cases}
\end{equation}
\subsection{Reduced system of ODEs corresponding to $U_4$ in Case 2.1}
The reduced system of ODEs corresponding to $U_4$ in Case 2.1 has the general form
\begin{equation}\label{lemeq1}
\begin{cases}
\frac{d^\al\varphi}{d z^\al} = a_1\psi+a_2 z\psi_z,\\
\frac{d^\al\psi}{d z^\al} = \varphi^{2m}(b_1\varphi+b_2 z\varphi_z),
\end{cases}
\end{equation}
where $a_1, a_2, b_1$ and $b_2$ are constants.
We formulate the following lemma with respect to a solution of the system given in (\ref{lemeq1}).
\begin{lem}
Let us assume that the parameter $m$ satisfies $m<0$ or $m>\frac{\al}{1-\al}$. Then, if the inequalities
\begin{equation*}
m\neq \frac{\alpha}{1-2\alpha},\quad
a_1-\frac{m+1}{m}a_2\al   \neq  0, \quad b_1-\frac{1}{m}b_2\al   \neq  0
\end{equation*}
hold, the system in (\ref{lemeq1}) has a solution of the form $\varphi(z)=c_1 z^{\lambda_1},$ $\psi(z)=c_2 z^{\lambda_2},$ where
\begin{eqnarray*}
\lambda_1&=&-\frac{\alpha}{m},\qquad \lambda_2=-\frac{(m+1)\alpha}{m},\\
c_1&=&\Biggl(\frac{\Gamma\left(1-\frac{\al}{m}\right)m^2}{\Gamma\left(1-\frac{(2m+1)\al}{m}\right)(ma_1-(m+1)a_2\alpha )} \frac{1}{(mb_1-b_2\alpha )}\Biggr)^{\frac{1}{2m}},\\
c_2&=&\frac{1}{\Gamma\left(1-\frac{(m+1)\al}{m}\right)}\Biggl(\frac{\Gamma\left(1-\frac{\alpha}{m}\right)^{2m+1}m^{2m+2}}{\Gamma\left(1-\frac{(2m+1)\al}{m}\right)}\frac{1}{(ma_1-(m+1)a_2\alpha )^{2m+1}(mb_1-b_2\alpha )}\Biggr)^{\frac{1}{2m}}.
\end{eqnarray*}
\end{lem}
\begin{pf}
Directly substituting $\varphi(z)=c_1 z^{\lambda_1}$ and $\psi(z)=c_2 z^{\lambda_2}$ into (\ref{lemeq1}) we obtain
\begin{equation}
\begin{cases}
c_1\frac{\Gamma(1+\lambda_1)}{\Gamma(1+\lambda_1-\al)}z^{\lambda_1-\al}=c_2(a_1 z^{\lambda_2}+a_2\lambda_2 z^{\lambda_2}),\\
c_2\frac{\Gamma(1+\lambda_2)}{\Gamma(1+\lambda_2-\al)}z^{\lambda_2-\al}=c_1^{2m+1}z^{2\lambda_1 m}(b_1 z^{\lambda_1}+b_2\lambda_1 z^{\lambda_1}).
\end{cases}
\end{equation}
The powers of $z$ appearing here should be equal in the two equations. From this observation, we have
\begin{equation*}
\lambda_1=-\frac{\al}{m},\qquad \lambda_2=-\frac{(m+1)\al}{m}.
\end{equation*}
Next, by the assumption of the lemma, we see that $\lambda_1>-1$ and $\lambda_2>-1$. From these results for $\lambda_1$ and $\lambda_2$,  $c_1$ and $c_2$ are obtained as in the statement of the lemma.
\qed
\end{pf}

The reduced system of ODEs corresponding to $U_4$ in Case 2.1 satisfies the conditions of Lemma 2, which implies that there exists a solution $(\varphi(z),\psi(z))$ as in the lemma. Then, from Tables 5 and 6, we obtain the following explicit invariant solution to (\ref{1}) with the condition that $m$ satisfies either $m<0$ or $m>\frac{\al}{1-\al}$:
\begin{equation}\label{19}
\left\{
\begin{array}{ll}
u(x,t)=&\left[\frac{m^2}{k^2(m+1)(2m+1)}\frac{\Gamma(-\frac{\al}{m})}{\Gamma(-\frac{(2m+1)\al}{m})}\right]^{\frac{1}{2m}}x^{\frac{1}{m}}t^{-\frac{\al}{m}},\\
v(x,t)=&\left[\frac{m^{2m+2}}{k^2(m+1)^{4m+1}(2m+1)}\frac{\Gamma\left(-\frac{\alpha}{m}\right)^{2m+1}}{\Gamma\left(-\frac{(m+1)\alpha}{m}\right)^{2m}\Gamma\left(-\frac{2m+1}{m}\alpha\right)}\right]^{\frac{1}{2m}} x^{\frac{m+1}{m}}t^{-\frac{(m+1)\al}{m}}.
\end{array}
\right.
\end{equation}
Finally, note that even though the reduced systems corresponding to $U_{2}$ in Case 2.1 and $U_3$ in Case 2.2 are of the form (\ref{lemeq1}), these systems do not satisfy the conditions of Lemma 2.
\subsection{Reduced system of ODEs corresponding to $U_5$ in Case 2.1}
By direct integration of the reduced system in this case, we obtain the following implicit solution with the condition $m<0$ or $m>\frac{\al}{1-\al}$: 
\begin{equation}\label{20}
\left\{\begin{array}{ll}
x=&\left(\frac{k^2}{(m+1)}\frac{\Gamma\left(1-\frac{(m+1)\al}{m}\right)^{2m}\Gamma\left(1-\frac{(2m+1)\al}{m}\right)}{\Gamma\left(1-\frac{\al}{m}\right)^{2m+1}}\right)^{\frac{1}{2m+2}}{\displaystyle\int_{\psi_0}^{\psi}\frac{d\theta}{\left(\theta^2+c_1\right)^{\frac{1}{2m+2}}}+c_2},\\
\varphi=&\left(\frac{m+1}{k^2}\frac{\Gamma\left(1-\frac{(m+1)\alpha}{m}\right)^2}{\Gamma\left(1-\frac{\alpha}{m}\right)\Gamma\left(1-\frac{(2m+1)\alpha}{m}\right)}\left(\psi^2+c_1\right)\right)^{\frac{1}{2m+2}}.
\end{array}\right.
\end{equation}
Here $\psi_0$ is an appropriately chosen lower bound, and $c_1$ and $c_2$ are constants. The invariant solution $(u(x,t),v(x,t))$ can be obtained from the above implicit solution and the form corresponding to  $U_5$ given in Table 5. Integrating the above solution explicitly is difficult, but for some particular values of the parameters $c_1,$ $c_2$ and $m$, explicit invariant solutions can be readily obtained. For example, for $c_1=0,$ we obtain the invariant solution
\begin{equation}\label{21}
\left\{
\begin{array}{lll}u(x,t) &=& \left(\frac{m^2}{k^2(m+1)(2m+1)}\frac{\Gamma\left(-\frac{\al}{m}\right)}{\Gamma\left(-\frac{(2m+1)\al}{m}\right)}\right)^{\frac{1}{2m}} (x-c_2)^{\frac{1}{m}}t^{-\frac{\al}{m}},\\
v(x,t)&=&\Biggl(\frac{m^{2m+2}}{k^2(m+1)^{4m+1}(2m+1)}\frac{1}{\Gamma\left(-\frac{(m+1)\al}{m}\right)^{2m}}\frac{\Gamma\left(-\frac{\al}{m}\right)^{2m+1}}{\Gamma\left(-\frac{(2m+1)\al}{m}\right)}\Biggr)^{\frac{1}{2m}}(x-c_2)^{\frac{m+1}{m}}t^{-\frac{(m+1)\al}{m}},
\end{array}
\right.
\end{equation}
which is identical to that obtained in \cite{jeff} for $m=\frac{1}{2}$ and $k=1$. Note that (\ref{19}) is invariant under the transformations corresponding to $X_2$ and $X_4$. This implies that (\ref{21}) is an invariant solution not only of the system of ODEs corresponding to $U_5$ but also of the system of ODEs corresponding to $U_4$ when $c_2=0$. Substituting $m=-\frac{1}{2}$ and $c_1\neq 0$ into (\ref{20}), we obtain another explicit invariant solution,
\begin{equation}\label{22}
\left\{\begin{array}{ll}
u(x,t)=&\frac{c_1}{2k^2}\frac{\Gamma\left(1+\alpha\right)^2}{\Gamma(1+2\alpha)}t^{2\al}\left[\tan^2\left(\frac{\sqrt{c_1}}{2k^2}\Gamma(1+\al)(x-c_2)\right)+1\right],\\
v(x,t)=&\sqrt{c_1}t^\al\tan\left(\frac{\sqrt{c_1}}{2k^2}\Gamma(1+\al)(x-c_2)\right).
\end{array}
\right.
\end{equation}
\subsection{Reduced system of ODEs corresponding to $U_2$ in Case 2.2}
It can be easily shown that the reduced system in this case has the solution
\begin{equation*}
\begin{cases}
\varphi(z) = a_1\frac{\Gamma(\alpha)}{\Gamma(2\alpha)}z^{2\alpha-1},\\
\psi(z) = cz^{\alpha-1}.
\end{cases}
\end{equation*}
Then, we obtain the following invariant solution of (\ref{1}) using the form of $U_2$ given in Table 9:
\begin{equation}
\begin{cases}
u(x,t)=a_1\frac{\Gamma(\alpha)}{\Gamma(2\alpha)}t^{2\alpha-1},\\
v(x,t)=a_2ct^{\alpha-1}+a_1xt^{\alpha-1},
\end{cases}
\end{equation}
where $c$ is a constant.
\subsection{Reduced system of ODEs  corresponding to $U_4$ in Case 2.2}
The solution to the reduced system in this case is
\begin{equation*}
\begin{cases}
\varphi(x)  =  \left(-\frac{a\Gamma(\al+1)}{k^2(1-2\al)}x+c_1\right)^{1-2\al},\\
\psi(x)  =  \frac{k^2(2\al-1)}{2a(1-\al)}\frac{\Gamma(2\al)}{\Gamma(\al)\Gamma(\al+1)}\left(\frac{a\Gamma(\al+1)}{k^2(2\al-1)}x+c_1\right)^{2(1-\al)}+c_2.
\end{cases}
\end{equation*}
With this and the form of $U_4$ given in Table 9, we obtain the following explicit invariant solution:
\begin{equation}
\left\{
\begin{array}{ll}
u(x,t) =& \left(-\frac{a\Gamma(\al+1)}{k^2(1-2\al)}x+c_1\right)^{1-2\al}t^{2\al-1},\\
v(x,t) =& \Biggl(\frac{k^2(2\al-1)}{2a(1-\al)}\frac{\Gamma(2\al)}{\Gamma(\al)\Gamma(\al+1)}\left(\frac{a\Gamma(\al+1)}{k^2(2\al-1)}x+c_1\right)^{2(1-\al)}-\al a \ln(t)+c_2\biggr)t^{\al-1}.
\end{array}
\right.
\end{equation}

We have thus found the group invariant solutions in all cases except for those corresponding to $U_{2}$ and $U_{3}$ in Case 2.1 and $U_{3}$ in Case 2.2. 
\section{Conclusion}
In this work, we have studied a class of nonlinear evolution systems of time fractional partial differential equations through application of Lie symmetry analysis. We presented a complete group classification of the system. Also, we derived one-dimensional optimal systems of Lie algebras of infinitesimal symmetries and reduced systems of fractional partial differential equations  to fractional and non-fractional systems of ordinary differential equations corresponding to the symmetries of the optimal systems in the case that the order of the partial differential equations, $\al$ satisfies $0<\al<1$. Furthermore, exact invariant solutions of the studied systems were obtained with the help of the reduced systems.
\section*{Acknowledgements}
We are very grateful to an anonymous referee for valuable suggestions and comments. We were able to improve the content of the work by addressing the points raised by the referee. This work was supported by JSPS (KAKENHI Grant No. 15H03613) and by the Science and Technology Foundation of Mongolia (Grant No. SSA-012/2016).


\section*{References}
\begin{thebibliography}{99}

\bibitem{Ovs} Ovsyannikov, L.V.: Group Analysis of Differential Equations. Academic Press, New York (1982)

\bibitem{blu1}
Bluman, G.W., Anco, S.C.: Symmetry and integration methods for differential equations. Applied Mathematical Sciences 154, Springer-Verlag, New York (2002)

\bibitem{blu2}
Bluman, G.W., Cheviakov, A.F., Anco, S.C.: Applications of symmetry methods to partial differential equations. Applied Mathematical Sciences, 168. Springer, New York (2010)

\bibitem{Ibr1} Ibragimov, N.H.: Elementary Lie Group Analysis and Ordinary Differential Equations. John Wiley, Chichester (1999)

\bibitem{Ibr2} Ibragimov, N.H.(Ed.): CRC Handbook of Lie Group Analysis of Differential Equations, Symmetries, Exact Solutions and Conservation Laws.  vol. 1. CRC Press, Boca Raton, Florida (1994)

\bibitem{ibra} Ibragimov, N.H.: A Practical Course in Differential Equations and Mathematical Modelling. Higher Education Press, World Scientific Publishing Co. Ptl. Ltd., Beijing (2010)

\bibitem{olver} Olver, P.J.: Applications of Lie Group to Differential Equations. Graduate texts in Mathematics, vol. 107, Springer-Verlag (2002)

\bibitem{Hydon} Hydon, P.E.: Symmetry methods for differential equations. A beginner's guide. Cambridge Texts in Applied Mathematics, Cambridge University Press (2000)

\bibitem{Gaz2} Gazizov, R.K., Kasatkin, A.A., Lukaschuk, Yu.S.: Continuous transformation groups of fractional differential equations. Vestnik UGATU 9, pp. 125--135 (2007) (in Russian)

\bibitem{Gaz3} Gazizov, R.K., Kasatkin, A.A., Lukashchuk, S.Yu.: Symmetry properties of fractional diffusion equations. Phys. Scr. (2009). doi:10.1088/0031-8949/2009/T136/014016

\bibitem{Sah} Sahadevan, R., Bakkyaraj, T.: Invariant analysis of time fractional generalized Burgers and Korteweg-de Vries equations. J. Math. Anal. Appl. 393, pp. 341--347 (2012)

\bibitem{Wang} Wang, G.W., Liu, X.Q, Zhang, Y.Y.: Lie symmetry analysis and explicit solutions of the time fractional generalized fifth-order KdV equation. Commun. Nonlinear Sci. Numer. Simul. 18, pp. 2321--2326 (2013)

\bibitem{Gaz1} Gazizov, R.K., Kasatkin, A.A.: Construction of exact solutions for fractional order differential equations by the invariant subspace method. Comput. Math. Appl. 66, pp. 576--584 (2013)

\bibitem{gupta} Gupta, A. K., Saha Ray, S.: On the Solutions of Fractional Burgers-Fisher and Generalized Fisher’s Equations Using Two Reliable Methods. International Journal of Mathematics and Mathematical Sciences (2014). doi:10.1155/2014/682910

\bibitem{Leo} Leo, R. A., Sicuro, G., Tempesta, P.: A foundational approach to the Lie theory for fractional order partial differential equations. Fract. Calc. Appl. Anal. 20, No 1, pp. 212--231 (2017)

\bibitem{huang} Huang, Q., Shen, S.: Lie symmetries and group classification of a class of time fractional evolution systems. J. Math. Phys. (2015). doi: 10.1063/1.4937755

\bibitem{jeff}  Jefferson, G.F., Carminati, J.: FracSym: Automated symbolic computation of Lie symmetries of fractional differential equations. Comput. Phys. Commun. 185, no. 3, pp. 430--441 (2014)

\bibitem{sing} Singla, K., Gupta, R.K.: On invariant analysis of some time fractional nonlinear systems of partial differential equations. I. J. Math. Phys. (2016). doi:10.1063/1.4964937

\bibitem{span} Oldham, K.B., Spaniel, J.: The Fractional Calculus. Academic Press, New York (1974)

\bibitem{Pod} Podlubny, I.: Fractional Differential Equations. Mathematics in Science and Engineering 198, Academic Press, San Diego (1999)

\bibitem{pat} Patera J., Winternitz. P.: Subalgebras of real three- and four- dimensional Lie algebras. J. Math. Phys., 18 (1977), 1449-1455.

\end{thebibliography}
\end{document}